\def\simlt{\lower.5ex\hbox{$\; \buildrel < \over \sim \;$}}
\def\simgt{\lower.5ex\hbox{$\; \buildrel > \over \sim \;$}}
\RequirePackage{lineno} 
 \documentclass[preprint2]{emulateapj}
 \usepackage{graphicx}

\newcommand{\myemail}{mrl@gps.caltech.edu}
\slugcomment{Accepted to the Astrophysical Journal }
\shorttitle{Uniform Retrieval Analysis}
\shortauthors{Line et al.}

\begin{document}
\title{A Systematic Retrieval Analysis of Secondary Eclipse Spectra II: A Uniform Analysis of Nine Planets and Their C to O Ratios}
\author{Michael R. Line}
\author{Heather Knutson}
\author{Aaron S. Wolf}
\author{Yuk L. Yung}
\affil{Division of Geological and Planetary Sciences, California Institute of Technology, Pasadena, CA 91125}
\email{mrl@gps.caltech.edu}
\altaffiltext{1}{Correspondence to be directed to \myemail}

\begin{abstract}
Secondary eclipse spectroscopy provides invaluable insights into the temperatures and compositions of exoplanetary atmospheres.  We carry out a systematic temperature and abundance retrieval analysis of nine exoplanets (HD189733b, HD209458b, HD149026b, GJ436b, WASP-12b, WASP-19b, WASP-43b, TrES-2b, and TrES-3b) observed in secondary eclipse using a combination of space- and ground-based facilities.  Our goal with this analysis is to provide a consistent set of temperatures and compositions from which self-consistent models can be compared and to probe the underlying processes that shape these atmospheres.  This paper is the second in a three part series of papers exploring the retrievability of temperatures and abundances from secondary eclipse spectra and the implications of these results for the chemistry of exoplanet atmospheres.  In this investigation we present a catalogue of temperatures and  abundances for H$_2$O, CH$_4$, CO, and CO$_2$.  We find that our temperatures and abundances are generally consistent with those of previous studies, although we do not find any statistically convincing evidence for super-solar C to O ratios (e.g., solar C/O falls in the 1-sigma confidence intervals in eight of the nine planets in our sample).  Furthermore, within our sample we find little evidence for thermal inversions over a wide range of effective temperatures (with the exception of HD209458b), consistent with previous investigations.  The lack of evidence for inversions for most planets in our sample over such a wide range of effective temperatures provides additional support for the hypothesis that TiO is unlikely to be the absorber responsible for the formation of these inversions.

\end{abstract}

\section{Introduction}
There are currently more than fiffty extrasolar planets with published secondary eclipse measurements (e.g., Knutson et al. 2010), with many more observations taken but not yet published.  These data come from both space and ground-based facilities, and span wavelengths ranging from the visible to the mid-infrared.  The combination of data from multiple telescopes spanning a broad range of wavelengths offers an invaluable tool for constraining the pressure-temperature profiles and compositions of exoplanetary atmospheres.  With an increasingly larger sample of spectra, we can begin to understand the underlying physical and chemical processes that control the atmospheric abundances through comparative exoplanetology.  

There are several interesting questions that we might address through a comparative study of exoplanet atmospheres.  One is the frequency of planets with super-solar C to O ratios.    The C to O ratio can potentially provide constraints on the region of the disk where the planet formed ({\"O}berg, Murray-Clay, \& Bergin 2011; Madhusudhan et al. 2011b).  These studies propose that planets that accrete their gas envelopes outside of the water snow line will have modestly super-stellar C to O ratios, which increase to even higher values for planets that form beyond the CO$_2$ ice line ({\"O}berg, Murray-Clay, \& Bergin 2011). 

 A uniform analysis of hot Jupiter atmospheres can also be used to investigate the origin of the temperature inversions detected in a subset of these planets.  It has been suggested (Hubeny et al. 2003; Fortney et al. 2008) that gas-phase TiO and VO, which are effective absorbers at optical wavelengths, could lead to the formation of temperature inversions in these atmospheres.  TiO and VO thermochemically exist in the gas phase in hotter planets ($T_{eq}>2000$ K).  Therefore, one would expect thermal inversions to be limited to planets that are hot enough to have gas phase TiO in their upper atmospheres. However, in a more recent study Spiegel et al. (2009) suggested that vigorous vertical mixing is required to keep both gas phase and condensed phase TiO and VO aloft in upper, inversion forming regions of the atmosphere.  Showman et al. (2009) and Parmentier, Showman \& Lian (2013)  demonstrated that TiO could be severely depleted due to cold traps in the deep atmosphere and on the planet's night side.  Furthermore, should TiO and VO persist despite the aforementioned reasons, Knutson et al. (2010) speculated that high amounts of UV flux from active stars might dissociate TiO and VO.   Recently Madhusudhan (2012) proposed that the abundances of TiO and VO could be significantly depleted if the atmosphere has a super-solar C to O ratio.  In this paper we use a uniform retrieval analysis to determine if there is a correlation between the presence or absence of a temperature inversion, the C to O ratio of the planet's atmosphere, and the activity level of its host star.

In Paper I (Line et al. 2013) we developed a suite of inverse modeling algorithms, called CHIMERA,  to determine the ranges of temperatures and compositions that were consistent with a given data set.  CHIMERA uses three Bayesian retrieval approaches including optimal estimation, bootstrap Monte Carlo, and differential evolution Markov chain Monte Carlo to determine the allowable ranges of temperatures and abundances for a given planet.  In this investigation we apply CHIMERA to nine planets in order to undertake the first uniform (meaning we use the same model, model assumptions, and retrieval approaches for each object) retrieval analysis of a set of exoplanet spectra.     Our uniform analysis allows us to make robust comparisons between planets, as we utilize the same model parameters and retrieval techniques for each target.  Comparing results derived from different retrieval approaches or different models can be complicated due to the differing model assumptions.    Such an analysis can also provide a useful set of statistical atmospheric properties from which detailed physical models such as general circulation models, photochemical models, and radiative equilibrium models can be compared.

Our goal in this study is to provide a catalogue of temperatures and abundances for nine well-observed planets and to address some of the outstanding questions regarding C to O ratios and the possible causes of thermal inversions.   Our sample includes the following planets: HD189733b, HD209458b, GJ436b, HD149026b, WASP-12b, WASP-19b, WASP-43b, TrES-2b, and TrES-3b.   These planets were selected because they span a wide range of physical parameters with a reasonably high number of spectral data points per planet.    In \S \ref{sec:Methods} we will describe the retrieval approaches and forward model assumptions.  The details of our approach can be found in Paper I.  In \S \ref{sec:Analysis} we present our temperature and abundance retrieval results for each of the nine planets and compare them to previously published analyses.  We then use the derived abundances to asses the allowed range of C to O ratios and comment on the implications of these results for current hypotheses for the origin of temperature inversions on these planets.   Finally, in \S\ref{sec:Conclusions} we will discuss the big picture view from our retrieval results, which show little evidence for C to O ratios larger than one.  
  
\section{Methods}\label{sec:Methods}
We summarize our retrieval methods and forward models here and refer the reader to Paper I for more detailed descriptions.   The goal of a retrieval, given some forward model, is to characterize the posterior probability distribution of the parameters of interest, in this case, temperatures and abundances via a radiative transfer model. This posterior is determined from a combination of prior information and the data.  Paper I describes three approaches that are commonly used to characterize posterior distributions. These include: optimal estimation (OE, e.g., Rodgers 2000), bootstrap Monte Carlo (BMC, Press 1992; Ford 2005), and differential evolution Markov chain Monte Carlo (DEMC, ter Braak 2006).   Optimal estimation minimizes a quadratic cost function using the Levenberg-Marquardt scheme and approximates the posterior as multivariate normal.  Bootstrap Monte Carlo uses a data resampling method based on a best-fit model to derive the parameter distributions.  Differential evolution Markov chain Monte Carlo is a type of Markov Chain Monte Carlo approach (MCMC) which uses a genetic algorithm to efficiently explore highly correlated parameter spaces.  DEMC and OE both evaluate the following log-likliehood function:
\begin{eqnarray} \label{eq:cost_func}
\chi^2({\bf x})={({\bf y}-{\bf F(x)})^{T}{\bf S_{e}^{-1}}({\bf y}-{\bf F(x)})} \nonumber \\
+({\bf x}-{\bf x_{a}})^{T}{\bf S_{a}^{-1}}({\bf x}-{\bf x_{a}}) 
\end{eqnarray}
where $\bf y$ is the set of $n$ data points, $\bf x$ is the $m$-dimensional parameter vector, $\bf F(x)$ is the forward model, and $\bf S_{e}$ is the $n \times n$ data error matrix.   $\bf x_{a}$ is the {\em a priori} state vector and $\bf S_{a}$ is the $m \times m $ {\em a priori} covariance matrix.    The first term in equation \ref{eq:cost_func} is simply the standard ``chi-squared" and the second term represents the prior knowledge of the parameter distribution before we make the observations.  

In Paper I we determined that the three approaches agree for high signal-to-noise, high-resolution data,  but diverge for the data sets currently available for typical hot Jupiters.  We also found that OE and DEMC tended to agree better than the BMC, and that our implementation of BMC was only able to characterize the probability distribution in the region very close to the nominal solution.  Therefore, in this investigation we utilize the OE and DEMC approaches to estimate the posterior probability distributions of the temperatures and abundances.   We use the results from OE to initialize the DEMC as described in Paper I. By using two distinct approaches, we can determine whether or not our results are sensitive to our choice of fitting method.  We present results from the DEMC approach in our final abundance catalogue, as this is the more widely accepted, robust approach for current exoplanet data (e.g, Benneke \& Seager 2012; Line et al. 2013).

The physical parameters we are most interested in are the temperature structure and the mole fractions of various gases.   We therefore choose to retrieve vertically uniform mixing ratios of H$_2$O, CH$_4$, CO, and CO$_2$.  These are generally the most thermochemically abundant (with the exception of CO$_2$) and infrared active species over the observational bandpasses for a variety of metallicities and C/O ratios.  Furthermore these species tend to have the most complete line lists either via experimentation or {\em ab initio} modeling (see Part I for description of our line lists).  We have assumed vertically uniform because the data do not provide useful constraints on the vertical abundance profile (Lee et al. 2012).  Also, the species of interest in this case, when abundant, thermochemically have near vertically uniform profiles to begin with, and vertical mixing tends to quench minor species resulting in vertically uniform profiles (Moses et al. 2011; Line et al. 2011).  We note that we have not included other potentially important absorbers such as NH$_3$\footnote{We have recently become aware that ExoMol group (Hill, Yurchenko, Tennyson (2013)) has released a new hot NH$_3$ cross-section data base.  We do not expect  the lack of NH$_3$ to have a significant impact on our retrievals because strong absorption is mainly limited to the 10 $\mu$m region and hence should not introduce strong correlations with the other absorbers. }, HCN, H$_2$S, C$_2$H$_2$, as accurate line-lists at high temperatures do not yet exist (to the best of our knowledge).  We plan to expand the scope of our future retrievals as more reliable sepctroscopic databases become available, and as improvements in the observations justify the inclusion of additional gases.

We also assume that these planets are hydrogen dominated and hence fix the H$_2$ and He mixing ratios to thermochemically appropriate abundances of 0.85 and 0.15, respectively, which hold over a wide range of temperatures and metallicities.   There could also be other optically inactive species such as N$_2$, O$_2$ , noble gases other than He, etc., but these molecules would have a minimal impact on the shape of our retrieved emission spectra and we therefore do not include them in our fits.  The retrievable species are assumed to have mole fractions much much less than that of H$_2$, therefore we do not include mean molecular weight effects on the spectrum.

We neglect to consider clouds in our models for the same reasons discussed Madhusudhan \& Seager (2009).  Furthermore,  Fortney et al. (2005) suggested that many cloud species (minor condensates and photochemical hazes) will have small normal optical depths resulting in minimal impact on emission spectra.  This idea is supported by recent findings that, though HD189733b (Pont et al. ()) and GJ436b (Knutson et al. (2014)) show strong evidence for high altitude absorbers in transmission, these planets still show strong signs of absorption in the emission spectra.   However, given the mounting evidence for clouds in primary eclipse, it is perhaps worth investigating the impacts of clouds on emission spectra in the future.

In Paper I we described two temperature retrieval approaches:  the level-by-level approach and the parametrized approach. For this analysis we choose to use the parametrized temperature profile approach. The limited number of data points per planet limits the practicality of a full level-by-level retrieval. The parameterization we use is based on analytic radiative equilibrium profiles (see Guillot 2010; Heng et al. 2012; Robinson \& Catling 2012) controlled by 5 free parameters.   These parameters are the infrared opacity, two visible opacities, partitioning of the two opacity streams, and a catch-all factor for the albedo, emissivity, and redistribution.  This last parameter effectively accounts for energy balance at the top of the atmosphere.  We do not try to self-consistently relate the temperatures back to the composition for reasons discussed in Paper I.       Once we have determine the uncertainty distributions for each of the five temperature profile parameters, we can then reconstruct the ensemble of temperature profiles.

Our priors are the same as those described in Paper I.  When using the optimal estimation formalism, by construction, the priors are Gaussian.  We choose very broad (12 orders-of-magnitude spanning the 68\% confidence interval) gas priors.  This mitigates the effect of the prior on the gas abundance retrievals.  We also choose broad Gaussian priors on the 5 parameters governing the parametrized temperature profile.  When using DEMC we choose flat gas priors, but use Gaussian priors on the temperature parameters.   These produce a spread in the reconstructed temperature profiles that is more consistent with numerical radiative equilibrium models than a flat prior would.  For any MCMC search when using flat priors one must impose limits in order to prevent the random walk process from venturing too far from useful phase space.  We choose a lower limit mixing ratio of 10$^{-12}$ and an upper limit of 0.1.  Although somewhat arbitrary, we would expect the molecular abundances of these four species to fall well within these limits.  

We obtain the planet  and system parameters (stellar radius, planet radius, stellar temperature, semimajor axis, planet gravity) from the published literature (see Table \ref{tab:Table0}).  These parameters are used in generating the parameterized temperature profile and when dividing by the stellar grid models.  We use interpolated (logg, Fe/H, and Teff) PHOENIX stellar grid models (Allard et al. 2000) to compute the contrast spectrum.  

We also should note that the retrieval results presented here must be taken in the context of our particular model.  Though not likely to change significantly, the retrieval abundances for the molecules presented here may change if we included a larger or different set of molecules.  It is worth investigating the impact of different molecules on our retrieval results through a nested model comparison in a future investigation, similar to the one presented in Swain et al. (2013). 

\section{Retrieval Analysis}\label{sec:Analysis}
In this section we provide a catalogue of abundances and temperatures for nine planets observed in secondary eclipse.  The sources of the secondary eclipse data are  shown in Table \ref{tab:Table1}.  We provide detailed descriptions of the results for each planet in our sample and compare these results to those of previous investigations.  Graphical results of the retrievals are summarized in Figures \ref{fig:Figure1}-\ref{fig:Figure3}.   Figure \ref{fig:Figure1} shows the secondary eclipse observations, the best fit spectra, and a statistical summary of all of the fits from the DEMC retrieval.  These fits are summarized with 68$\%$ and 95$\%$ confidence bounds along with a median spectrum.   Figure \ref{fig:Figure2} shows the temperature profiles summarized with 68$\%$ and 95$\%$ confidence bounds, a median profile, and a best fit.  We also show the averaged thermal emission contribution function to get a sense for which pressure levels the observations probe.  For comparative curiosity we show, in some cases, self-consistent first-principle temperature profiles from the literature.  Finally, Figure \ref{fig:Figure3} shows the marginalized gas abundance posteriors for each planet along with the imposed priors (flat for DEMC, Gaussian for OE).  We use the resulting retrieval results to derive the C to O ratio probability distributions in Figure \ref{fig:Figure4}.    These distributions generally have a double-peaked structure, which is the manifestation of the uniform abundance priors in the C to O distribution (see Paper I for a detailed discussion of this issue).  Rather than showing the two peaked C to O distribution resulting from the prior and posterior together, we normalize the posterior derived C/O distribution by the double-peaked prior C/O distribution to give us a sense for how the data contributed to our knowledge of the C to O ratio.  This has no formal statistical meaning, but it provides a clear visual representation of the information provided by the observations independent of the assumed priors.

 We also include a Table (Table \ref{tab:Table2}) comparing our numerical results to those of previous studies for easy reference. We provide our 68$\%$ confidence intervals for the molecular mixing ratios along with the nominal best-fit values derived from DEMC.  We also report the reduced cost function value from equation \ref{eq:cost_func}, $\chi^2/N$ for the best fit, where N is the number of data points.  A $\chi^2/N$of one suggests that the model on average fits the data within the 1-sigma error bars (just as in Madhusudhan \& Seager 2009). Since we place limits on the flat gas priors used in the DEMC retrieval, we must interpret the retrieved abundance range in the context of those limits.  Given those limits, the 68$\%$ confidence interval from the flat gas prior would result in abundances that span $8.15\times10^{-11} - 1.51\times10^{-2}$, or $\sim$8 orders of magnitude.  Anything smaller than this suggests that the data was informative within the context of our model.  Furthermore, we quote the 68$\%$ confidence interval for the C/O ratios derived from our abundance retrievals.  These values are computed directly from the posterior C to O distributions, not the normalized-by-prior distribution shown in Figure \ref{fig:Figure4}. Given the imposed abundance limits,  the 68$\%$ confidence interval in the C/O distribution resulting from the uniform gas priors  is $5.10\times10^{-2} - 1.45\times10^{1}$, or  $\sim$ 3 orders of magnitude.

\subsection{HD189733b}
Together with HD209458b, HD189733b is one of the best-studied hot Jupiters to date.  This is because it orbits a bright, nearby K star with a favorable planet-star radius ratio.  Secondary eclipse observations have been made with a variety of instruments including HST-NICMOS (Swain et al. 2009), Spitzer IRAC (Deming et al. 2006;  Charbonneau et al. 2008; Knutson et al. 2009; 2012; Agol et al. 2012), MIPS (Knutson et al. 2009; Charbonneau et al. 2008), and IRS (Grillmair et al. 2007; 2008) spanning $\sim$1.5 $\mu$m to $\sim$25 $\mu$m.  Several previous retrieval investigations have constrained the range of allowable temperature structures and compositions for this atmosphere.  The first complete study via a systematic parametric grid search was published by Madhusudhan \& Seager (2009).  The composition they derived was high (relative to a solar composition atmosphere) in CO$_2$ and CO followed by low abundances of CH$_4$ and a moderate abundance of H$_2$O.  Lee et al. (2012) and Line et al. (2012) came to similar conclusions using the optimal estimation retrieval approach.  The large abundance of CO$_2$ relative to the other species remains chemically perplexing  (Line et al. 2010;  Moses et al. 2011;  2013a).  There is also no evidence for a thermal inversion. 

Our abundances are generally consistent with the previous results to within an order of magnitude.  However, not all of the previous investigations are consistent.  For instance, all of the results for HD189733b in Table \ref{tab:Table2}, including ours, require an anomalously high abundance of CO$_2$ aside from the analysis presented in Swain et al. (2009) (see Shabram et al. (2011) for a more detailed discussion of this discrepancy).  Also, the upper limit on CH$_4$ quoted by Line et al. (2012) is much higher than the limits from previous studies.  This is likely due to the problems associated with using optimal estimation in data regimes in which the resulting posterior probability distributions are non-Gaussian.   We note that the data used in the retrievals were not the same in all cases.  We simultaneously retrieved the abundances and temperatures using a combination of data including the NICMOS, IRS, and Spitzer photometry.  Madhusudhan \& Seager (2009) considered these data sets separately. Swain et al. (2009) and Line et al. (2012) only considered the NICMOS data set.  Lee et al. (2012) considered all available data sets.  It is worth considering whether individual data sets give consistent results, as data taken at different epochs could vary as a result of stellar variability, differences in detector systematics, etc.  Despite these differences the results of all of these studies are in generally good agreement, suggesting that our conclusions for this planet are reasonably robust.

 In our new study we find an enhanced best fit C/O ratio of 0.85 (0.47-0.90), well within the range quoted by previous investigations.  This is a fairly robust result relative to the prior (see Figure \ref{fig:Figure4}).  
 
The large number of Spitzer IRS data points that have thermal emission weighting functions near the 10 mbar level combined with our requirement of radiative equilibrium via the analytic parameterization, provide a well constrained temperature profile above the 0.1 bar level.     Below this level the temperature profile begins to diverge because very little emission is able to escape from these deeper regions of the atmosphere.   Our derived spread in profiles in the well-constrained regime between 10-100 mbar is in agreement with the profile derived in Lee et al. (2012).  We also show a temperature profile from Moses et al. (2011), which is generally cooler than  our derived spread.

\subsection{HD209458b}\label{sec:HD209}
HD209458b is also a favorable target for secondary eclipse observations due to its relatively deep eclipse depths and bright host star.  Multiple groups have obtained secondary eclipse observations using broadband photometry (Cowan et al. 2007; Knutson et al. 2008; Crossfield et al. 2012a),  NICMOS spectroscopy (Swain et al. 2009b), and Spitzer IRS spectroscopy (Richardson et al. 2007; Swain et al. 2008).  The consensus from these observations is that this planet possesses a strong thermal inversion (Burrows et al. 2007; Madhusudhan \& Seager 2009;  Swain et al. 2009) and poor day-night heat redistribution (Cowan et al. 2007).   Madhusudhan \& Seager (2009) used Spitzer photometry to constrain the range of allowable dayside temperature profiles and compositions.  Their ensemble of fits suggest a strong thermal inversion beginning near the $\sim$500 mbar level and a high C/O ratio.  However, they point out that simple 1-dimensional dayside averaged profiles may result in over-interpretation of the data.  Combining NICMOS  and IRS spectroscopy with Spitzer broadband photometry, Swain et al. (2009) salso found that a thermal inversion is needed to explain the combined observations.  They also placed some limited constraints on the molecular abundances.

Our independent analysis focuses on the broadband data only.  We are unable to simultaneously fit the NICMOS/IRS spectroscopy and the Spitzer photometry with a single model that provides a good match to all of the observations.  Our best fits including both data sets generally result in reduced $\chi^2$ values greater than 5, and hence such solutions may not be statistically meaningful.  Our broadband retrieval confirms the existence of a strong (median inversion depth-to-error at the temperature minimum of 7.5, or 7.5$\sigma$) thermal inversion with the temperature minimum occurring within the  1bar -100 mbar region.  This tropopause location is consistent with tropopause locations in our own solar system and with what is predicted by Robinson \& Catling (2013), but is somewhat deeper than what is shown in Moses et al. (2011).   We also find C to O ratios near unity.  The inversion is needed to explain the high 4.5 and 5.7 $\mu$m fluxes relative to the other points.  This is also the location of the strong CO band.  Large abundances of CO ($>5\times10^{-4}$) are required to push the weighting functions at these wavelengths high up in the atmosphere where the inversion is strongest.  This high abundance of CO dominates the C to O ratio forcing C/O's of unity, again consistent with the results of Madhusudhan \& Seager (2009).  

The temperatures at pressure levels higher than the $\sim$50 mbar level exceed those of the TiO condensation curve supporting the hypothesis that TiO is the cause for inversions.  However, these temperatures exceed the TiO condensation curve {\em because} of the inversion.   If the inversion were not present, the temperatures at these levels would be cooler than the TiO condensation curve, begging the question of whether or not TiO is actually the cause of the inversion or if it is some other absorber.  We suppose though, that if the planet had a higher internal heat flux when it first formed,  it may have reached this state by cooling from a hotter profile where TiO and VO were in gas phase everywhere

Future observations with the HST WFC3 may should provide additional confirmation of the existence of a temperature inversion.  The CO feature near 1.6 $\mu$m probes altitude regions near the 100 mbar level--where the temperature is increasing from the minimum--whereas the bluer wavelengths probe the few bars levels. Hence these wavelengths probe regions of the atmosphere that bracket the temperature minimum, as determined by the broadband data.    The existence of this feature will confirm the presence of a temperature inversion as well as the high CO abundance.

\subsection{GJ436b}
GJ436b is a warm ($\sim$700-900 K) Neptune-mass planet with a seemingly unusual atmosphere chemistry that has generated considerable attention in the modeling community.  The first set of Spitzer observations (Stevenson et al. 2010) indicated that this planet is rich in CO and depleted in CH$_4$.  This conclusion is primarily driven by the high 3.6 to 4.5 $\mu$m flux ratio.  This composition is at odds with its temperature given the assumption of thermochemical equilibrium and solar elemental abundances.  One would expect such a planet to be rich in methane and low in CO.  These authors suggested that photochemistry might be responsible for the apparent depletion of methane.  However, photochemical depletion of methane does not appear to be significant in hydrogen dominated atmospheres in this temperature range even under a wide variety of assumed vertical mixing strengths and UV fluxes (Line et al. 2011; Miller-Ricci et al. 2012; Moses et al. 2013b).  It has recently been suggested that this planet has an extremely metal-rich (300-2000$\times$ solar) atmosphere, which provides a simple explanation for the measured enhancement of CO over CH$_4$ without the need to invoke extreme or exotic chemistry (Moses et al. 2013b).  

Table \ref{tab:Table2} shows how our results compare to those of Stevenson et al. (2011) and Madhusudhan \& Seager (2011).  We also conclude that the atmosphere is rich in CO and depleted in CH$_4$ relative to what one would expect for a solar composition atmosphere at these temperatures.  We point out that although we quote a confidence interval, on all four gases, our fits only provide upper limits on most species with the exception of CO.  Our lower limit for these unconstrained gases comes from our prior, which artificially places a hard limit at $1\times10^{-12}$.  If we examine the histogram for methane in Figure \ref{fig:Figure3} we see that an appropriate upper limit might be placed at the half max location of $\sim10^{-7}$, which is within an order of magnitude of the previous results. We also find upper limits on the CO$_2$ and H$_2$O abundances, but again no strong lower limit.  The CO$_2$ marginalized posterior shows a multimodal behavior, with a strong mode occurring at $\sim10^{-7}$ and a weaker mode near  $\sim10^{-3}$.   Our estimate of the C to O ratio is similar to that quoted by Madhusudhan \& Seager (2011) as the most probable values seem to fall between solar and unity.  Figure \ref{fig:Figure4} indicates that a C to O ratio greater than one is highly improbable. Our temperature profile dispersion is nearly identical to those found in Madhusudhan \& Seager (2011) between $\sim$0.1-1 bars and is generally hotter than the Lewis et al. (2010) derived general circulation model profile.

\subsection{HD149026b}
To date there has been no detailed atmospheric retrieval analysis for this planet.  It is a Saturn-mass planet with a very large core orbiting a metal-rich star (Sato et al. 2005), and therefore is a good candidate for a metal-rich atmosphere.  Furthermore, its likely high atmospheric metallicity and temperature make it a prime candidate for an inversion caused by gas-phase TiO and VO (Fortney et al. 2006).  Stevenson et al. (2012) obtained Spitzer photometry and interpreted the observations using the self-consistent models from Fortney et al. (2005; 2006; 2008).  These data-model comparisons suggest that the planet's emission spectrum is well-described by a relatively high (30$\times$ solar) metallicity atmosphere with correspondingly enhanced CO and CO$_2$ features, with no evidence for a temperature inversion.

 Our retrieval results indicate that the planet has more CO and CO$_2$ than methane, consistent with the Stevenson et al. (2012) results.  The marginalized gas posteriors (Figure \ref{fig:Figure3}) show an upper limit on methane, a strong peak in the probability distribution for water near $\sim10^{-5}$ with an unconstrained tail towards low abundances, and a preference for large ($>10^{-4}$) abundances of CO.   These results are consistent with the planet's relatively high atmospheric temperature ($\sim$1700 K), which tends to favor CO over CH$_4$ at near-solar abundances.  There is a slight preference for high abundances of CO$_2$ (relative to solar) consistent with the Stevenson et al. (2012) results.  We find a C to O ratio that is consistent with solar and can rule out ratios greater than unity.  Our temperature profiles are most consistent with the solar metallicity model from Stevenson et al. (2012) without an inversion.  Our range of temperature profiles are also similar to those in Fortney et al. (2006) over the IR photosphere.  We are in good agreement with their general conclusions that this planet lacks a strong dayside temperature inversion.  
  
\subsection{WASP-12b}\label{sec:WASP12b}
The atmosphere of hot Jupiter, WASP-12b has generated some excitement as the first candidate for a planet with a super-solar C to O ratio.  Madhusudhan et al. (2011a) carried out a retrieval analysis and found that their fits preferred a high C/O ratio atmosphere based on seven broadband photometry points ranging from J-band through the Spitzer IRAC 8 micron point.  This conclusion stems primarily from the apparent lack of water absorption in the near-infrared.  Due to the difficulty in removing water via disequilibrium processes, Madhusudhan et al. (2011a) argue that the most plausible explanation for the low water abundance is a high C to O ratio.    {\"O}berg et al. (2011) have suggested that gas giant planets could form with different C to O ratios depending on their location in the protoplanetary disk, suggesting that C to O ratios for hot Jupiters could provide constraints on their formation locations.  

Since the publication of the Madhusudhan et al. (2011a) analysis there have also been some corrections to the estimated secondary eclipse depths for this planet.  Bergfors et al. (2013) identified a companion star to WASP-12 responsible for contaminating the measured photometry.  Crossfield et al. (2012) reanalyzed the Spitzer data to account for contamination from the binary M star companion (Bechter et al. 2013), which dilutes the light from the primary star and decreases the measured eclipse depths.  The same paper also presents new K band measurements along with the HST WFC3 data from Swain et al. (2013). Their updated data-model comparison indicates that a nearly isothermal atmosphere without the presence of significant absorbers could explain the data.    Swain et al. (2013) observed WASP-12b in both emission and transmission using the HST WFC3.  Like Crossfield et al. (2012), they also find that the data do not require a high C to O ratio, particularly if oxygen bearing species such as TiO and VO are included to explain the transmission spectrum.

In our analysis we use the most current data that could be found in the literature.  This includes the Subaru narrowband photometric 2.315 $\mu$m point from Crossfield et al. (2012b), an updated K$_s$ band photometric point from Zhao et al. (2012) (larger error bar than the previous K$_s$ band data from Croll et al. 2012) and we account for the ``null" and ``ellipsoidal" 4.5 $\mu$m analysis in Cowan et al. (2012).  All of the data have been updated to account for the dilution correction factor discussed and implemented in Crossfield et al. (2012) and Swain et al. (2013).  

We applied several different analyses.  First, we retrieve the temperatures and abundances under the assumption of the  Cowan et al. (2012) ``null" hypothesis for the 4.5 $\mu$m flux.  Crossfield et al. (2012b) suggests (via private comm. from N. Cowan) that the flux resulting from this analysis is likely more reliable than the flux resulting from the ``ellipsoidal" analysis due to the more realistic assumptions about the planetary shape.  Hence, we choose the retrieval results that derive from the set of observations that include the 4.5 $\mu$m ``null" data point as our nominal result.  We find a nearly isothermal atmosphere with the potential for a weak inversion ( $\sim$2.5$\sigma$ as defined in \S \ref{sec:HD209}), though the small temperature gradients near the temperature minimum are enough to invoke absorption and emission features (Figure \ref{fig:Figure1}).   The self-consistent temperature profile from Crossfield et al. (2012b) is in very good agreement with our retrieved spread.  

The preferred solutions from the DEMC favor an atmosphere that is very abundant in CO$_2$ (Figure \ref{fig:Figure3}, blue histogram).  This high abundance of CO$_2$ drives the C to O ratio closer to solar than to unity.  If we repeat the results using the 4.5$\mu$m ``ellipsoidal" derived data point, the lower flux value drives the CO$_2$ abundance to even higher levels.  The inversion depth strengthens slightly.  In both cases we note that the water abundance can span a wide range.  This is different than in Madhusudhan et al. (2011) where in general the retrieved water abundances were lower than ~10 parts-per-million, which was the primary driver in determining a high C/O.  Our gas abundance retrievals do not provide strong evidence for a high C/O atmosphere, though we cannot rule out such a possibility  (blue and red curves in Figure \ref{fig:Figure4}).

One might question how realistic solutions with such high abundances of CO$_2$ may be. Such high CO$_2$ abundances are generally not thermochemically permissible in highly reducing hot-Jupiter-like atmospheres.    In order to explore this we have placed an upper limit on the CO$_2$ abundance at a mixing ratio of 10$^{-5}$ and repeated the retrievals under the ``null" and ``ellipsoidal" assumptions.    We show the resulting gas posterior probability distributions as the gray curves in Figure \ref{fig:Figure3}.  The upper limit prevents the DEMC from finding the high CO$_2$ abundance mode.  The other molecules then compensate to fit the data. The best fit is not as good ($\chi^2_{best}/N\sim$2, not shown in Table \ref{tab:Table2}) as if we permit the higher CO$_2$ abundances, though not by a statistically significant amount.  Imposing this limit has virtually no effect on the CH4 and CO (which is completely unconstrained as indicated by its nearly flat posterior in Figure \ref{fig:Figure3}) abundances.  The imposed upper limit does, however, drive up the water abundance.  Again, we are unable to find strong evidence for C to O ratios larger than unity (Figure \ref{fig:Figure4}, gray curves).  The imposed upper limit on the CO$_2$ abundance has little effect on the temperature profile--no strong evidence for an inversion. 

The lack of a strong inversion ($\sim$2.5$\sigma$ ) suggests that perhaps TiO and VO are not strongly absorbing, though a nearly isothermal atmosphere suggests that perhaps there is some high altitude opacity.   In a high C/O atmosphere, thermochemically, TiO and VO should be depleted as suggested by Madhusudhan (2012).  However, we do not find strong evidence for a high C/O atmosphere.  

It is difficult to disentangle the reason for the discrepancy between our results and those of Madhusudhan et al. (2011).  Potential differences could be due to the use of different data sets (dilution corrected fluxes vs. non corrected) and different model assumptions such as the assumed gases, gas opacity databases, or the temperature profile parameterization.  A detailed model comparison over a wide range of parameter space would be needed to identify potential differences.   Future observations and modeling of this intriguing planet are certainly needed in order to draw firm conclusions.

\subsection{WASP-19b}
 Given its high equilibrium temperature (2400 K) and the assumption of solar elemental abundances, WASP-19b is expected to have moderate quantities of TiO and VO, hence causing a thermal inversion (Fortney et al. 2008).  Anderson et al. (2013) and Madhusudhan (2012) found no evidence for an inversion, leading Madhusudhan (2012) to suggest a high C to O ratio in order to deplete TiO and VO.  WASP-19 is one of the most active stars to host a hot Jupiter, therefore UV destruction of an absorbing molecule could also provide an alternative explanation (e.g. Knutson et al. 2010).  Madhusudhan (2012) explored both carbon-rich and oxygen-rich models for this planet and concluded that the observations are unable to constrain the C to O ratio, but may slightly favor a C to O ratio greater than one.  

Figure \ref{fig:Figure3} suggests that CO and methane are almost completely unconstrained, as their marginalized posteriors closely track the prior.  The data do appear to place an upper limit on water ($\sim10^{-3}$) and a weak upper limit on CO$_2$ ($\sim10^{-2}$). We concur with Madhusudhan (2012), who conclude that the current data for this planet do not provide useful constraints on the C to O ratio (Figure \ref{fig:Figure4}).  Both the oxygen rich and carbon rich derived abundances of Madhusudhan (2012) are consistent with our 68$\%$ confidence intervals as well as the temperature profiles.  Future observations of WASP-19b with WFC3 would greatly reduce the uncertainties in the gas abundances and temperature profile.  Figure \ref{fig:Figure1} shows a large divergence in the spectra at wavelengths less than 2 $\mu$m.  WFC3 observations with a signal to noise similar to those obtained for WASP-12b would narrow this dispersion down and ultimately reduce the uncertainties in the abundance of H$_2$O. A strong lower bound on the water abundance would provide improved constraints on the C to O ratio.

\subsection{WASP-43b}
WASP-43b has a very favorable planet-star radius ratio and correspondingly deep secondary eclipses, making it a prime target for both ground and space-based observations.  It is currently one of the coolest ($\sim$1500 K) planets accessible via ground based observations.  Blecic et al. (2013) explored the composition of WASP-43b using the temperature and abundance retrieval approach of Madhusudhan et al. (2011a).  They found that the composition is only weakly constrained, but rule out a temperature inversion based on the relative fluxes between the Spitzer IRAC channels and the ground based J and K band photometry.  We obtain stronger constraints from our fits that allow us to confidently rule out C to O ratios larger than one (Figure \ref{fig:Figure4}). This is primarily due to the upper limit on methane near $\sim10^{-5}$ (Figure \ref{fig:Figure3}).     We also find that the temperature profile lacks a temperature inversion, consistent with the conclusions of Blecic et al. (2013).  

\subsection{TrES-2b}
TrES-2b is a highly irradiated hot-Jupiter that is theoretically predicted to possess a temperature inversion (a ``pM" class planet, Fortney et al. 2008).  Croll et al. (2010a) compared the spectral energy distribution of photometric data points at 2.14, 3.6, 4.5, 5.6, and 7.8 $\mu$m to forward models of Fortney et al. (2005; 2006; 2008) which assume local thermal equilibrium and solar metallicity.  Depending on the details of the temperature profile,  TiO could be present in gas phase, leading to the formation of a temperature inversion in TrES-2b's atmopshere.  They found that cooler temperature profiles fit the data better, naturally explaining the absence of an inversion as TiO and VO should condense out of the planet's dayside atmosphere.

We find that in our fits the data provide minimal constraints on the relative molecular abundances.  The marginalized posterior distributions show very little preference for a specific combination of abundances (i.e., there is no strong mode, Figure \ref{fig:Figure3}).   Although we quote a 68$\%$ confidence interval for each species, we note that these confidence intervals are nearly the same as those given by the prior, suggesting relatively weak constraints from the data aside from a slight preference for higher abundances of methane.  As a result of these weak constraints, we cannot make strong statements about the C to O ratio.  Our retrievals rule out a thermal inversion, in agreement with the results from Croll et al. (2010a).  We find that the dispersion in the temperature profiles straddles the TiO condensation boundary, with temperatures between 300 mbars and 0.7 mbars dipping below the condensation temperatures.  This suggests that TiO and VO would likely be lost to cold traps in these cooler regions even if there are local regions in the atmopshere that are warm enough for them to exist in gas phase (e.g., Spiegel et al. 2009; Showman et al. 2009; Parmentier, Showman \& Lian 2013).

\subsection{TrES-3b}
Like TrES-2b,  TrES-3b is an interesting target for identifying thermal inversions that may be due to TiO (Fortney et al. 2008).  Croll et al. (2010b) examined H and K band photometry combined with Spitzer photometry in order to determine the temperature structure.  They found that the atmosphere could be explained with an isothermal temperature structure.      Our retrievals allow us to expand on these initial conclusions, which were based on a comparison to the same class of forward models by Fortney et al. cited in the previous discussion on TrES-2b.  We find that H$_2$O is fairly well constrained with abundances near $10^{-4}$.  This is reasonable for a planet at these temperatures with solar composition.  Methane has a well defined upper limit of $\sim10^{-6}$, again consistent with solar composition.  CO remains unconstrained due to the large uncertainty on the IRAC 4.5 $\mu$m data point and the lack of constraints from data at other wavelengths (Figure \ref{fig:Figure3}).  CO$_2$ has a weak upper limit near $10^{-4}$.  This upper limit arises from the wings of the 2.1 $\mu$m CO$_2$ band probed by the K band photometry.   These abundances are generally consistent with a solar composition atmosphere.  We can also confidently rule out C/O ratios larger than one due to the well constrained water abundance and low upper limit to the methane abundance.  The temperature profile dispersion for TrES-3b is largely cooler than those of TrES-2b over the range of pressures probed by these infrared observations, and lies well below the condensation curve for TiO.  The lack of inversions in TrES-2 and -3b therefore appears to be consistent with the hypothesis that TiO and VO could be the inversion-causing opacity sources.

\section{Discussion \& Conclusions}\label{sec:Conclusions}
Secondary eclipse spectra simultaneously tell us about the dayside temperature structures and compositions of the planets examined in our study.  We have performed a comprehensive, uniform retrieval analysis of nine exoplanets observed in secondary eclipse and provide a catalogue of the resulting temperature and molecular abundance estimates.    Such analyses provide a useful set of atmospheric information that can be used to test various hypotheses about the origin of the observed properties of hot Jupiter atmospheres.  Our results are generally consistent with those of previous investigations despite differences in model assumptions and retrieval approaches.  However, unlike previous investigations, we find no statistically compelling evidence in support of C to O ratios larger than solar as shown in Figure \ref{fig:Figure4} (though HD209458b appears to have a C/O ratio near unity due to the retrieved CO abundance).   Solar C/O falls within the 68\% confidence interval for eight of the nine planets in our sample.   We also find no strong evidence for thermal inversions eight of the nine planets we investigated, with the exception being HD209458b.  WASP12b has a nearly isothermal atmosphere, but there is not strong evidence for an actual inversion.  Out of our sample of nine planets, three (HD209458b, WASP-19b, WASP-12b) are hot enough to contain TiO in the gas phase, but our retrievals do not strongly detect inversions in two of their atmospheres.  Madhusudhan (2012) have suggested that high C to O ratios could reduce the gas phase TiO and VO abundances thus eliminating the absorbers that cause inversions.  However,  we do not find strong evidence in support of high C to O ratios for the two planets hot enough to posses gas phase TiO and VO, although we cannot rule out the possibility high C to O ratios in WASP-19b and WASP-12b.  We do however find an inversion in the atmosphere of HD209458b while also having a high C to O ratio--contradictory to the idea proposed by Madhusudhan (2012), if TiO and VO are indeed the responsible absorbers.  This would seem to provide another line of evidence against TiO and VO as the absorber responsible for the inversions, although it does not rule out the more general hypothesis that variations in elemental abundances are responsible for the lack of temperature inversions in a subset of hot Jupiters.  

\begin{figure*}[h]
\begin{center}
\includegraphics[width=0.95\textwidth, angle=0]{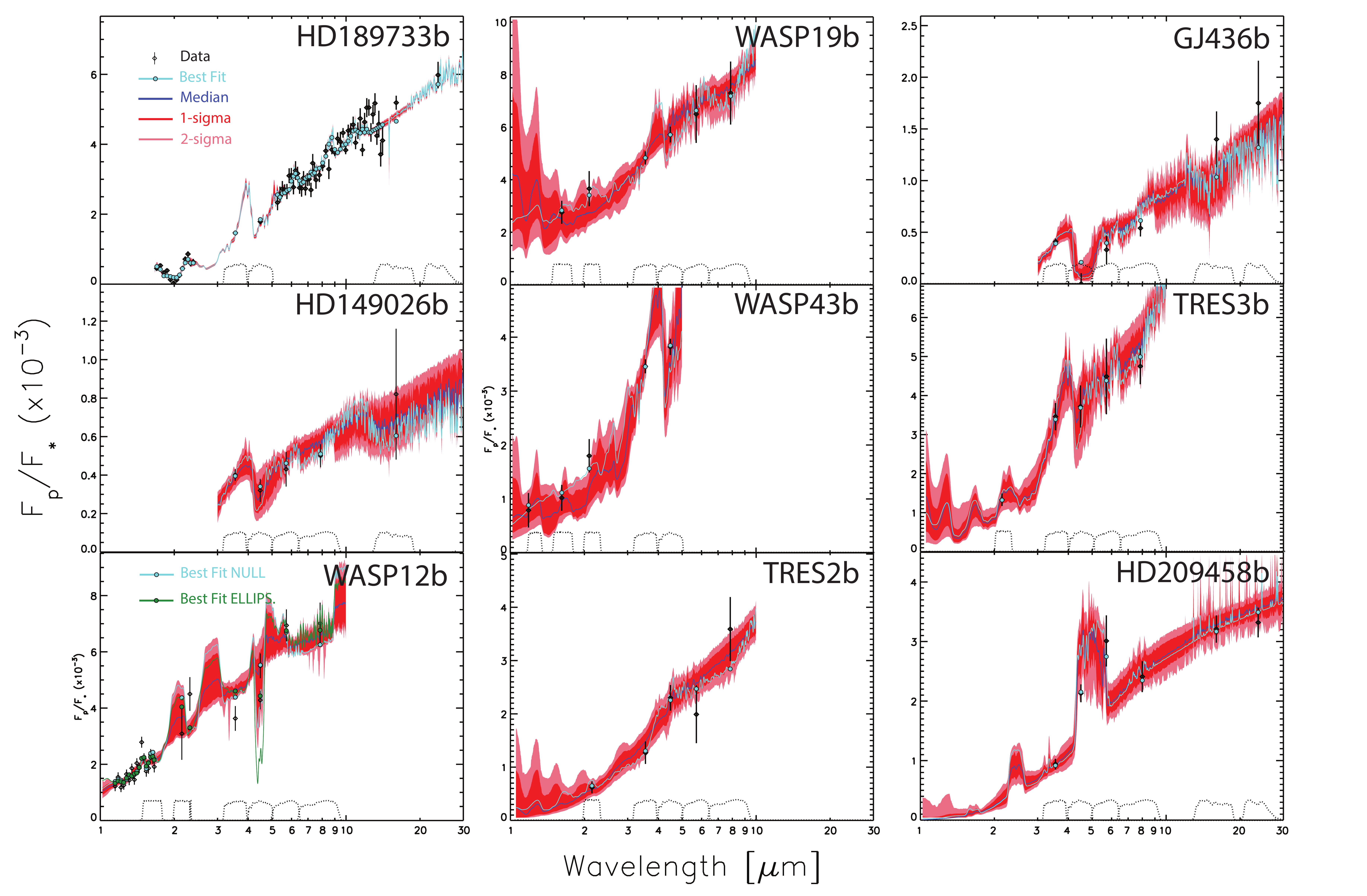}
\end{center}
     \caption{ \label{fig:Figure1} Secondary eclipse spectra and fits for each planet resulting from the differential evolution Markov chain Monte Carlo retrieval approach.  The diamonds with error bars are the data from the sources in Table \ref{tab:Table1}. The DEMC retrievals produce several hundred thousand spectra.  The best fit of this ensemble is shown in light blue.  The light blue circles are the best fit model binned to the data.  The ensemble of spectra are summarized with the median spectra (blue) and the 1- and 2-sigma confidence intervals (dark red and light red respectively).  The green spectrum in the WASP-12b panel is the best-fit spectrum when including the ``ellipsoidal" variations to the 4.5 $\mu$m datum (see \S\ref{sec:WASP12b}).  The dotted curves at the bottom of each panel are the filter profiles for each broadband measurement. }
\end{figure*}

\begin{figure*}[h]
\begin{center}
\includegraphics[width=0.8\textwidth, angle=0]{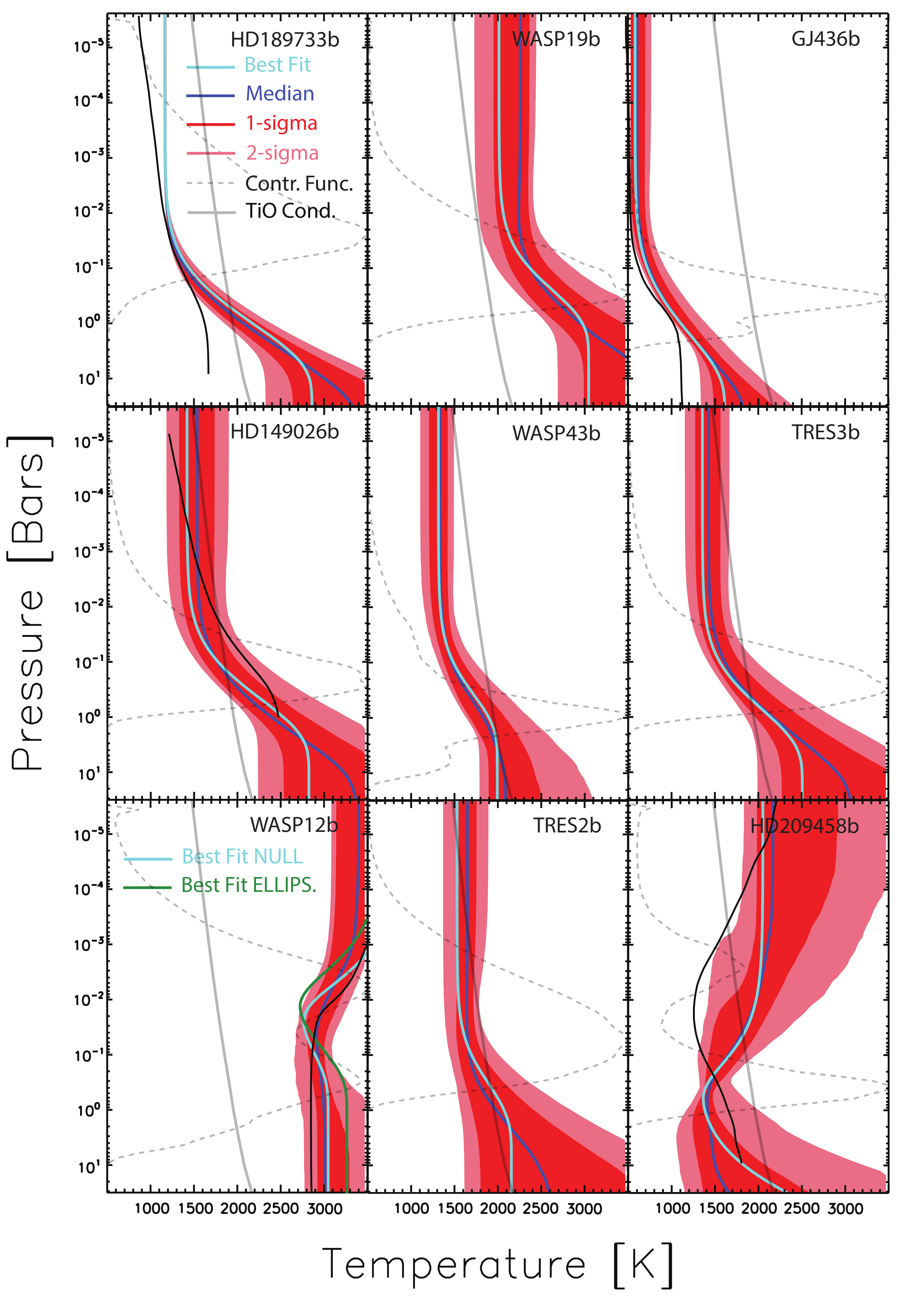}
\end{center}
     \caption[Mixing Ratio Vertical Profiles]{ \label{fig:Figure2} Summary of the retrieved temperature profiles.  The best fit temperature profile is shown in light blue.   The ensemble of temperature profiles are summarized with the 1 (dark red)- and 2 (light red)-sigma confidence intervals and the median of the profiles (dark blue).  The light gray dashed curves are the wavelength averaged thermal emission contribution functions.  These represent on a whole, where the emission is coming from.  The solid light gray line in each panel is the TiO condensation curve for 1x solar composition (Fortney et al. 2008). The green profile in the WASP-12b panel is the best-fit temperature profile when including the ``ellipsoidal" variations to the 4.5 $\mu$m datum (see \S\ref{sec:WASP12b}).  The black profiles in some of the panels are temperature profiles derived from from first principles from the following sources: HD189733b-Moses et al. 2011 ``1D $2\pi$",     GJ436b-Lewis et al. (2010) 1$\times$ solar , HD149026b-Stevenson et al. (2012) 1$\times$ solar, WASP12b-Crossfield et al. (2012), HD209458b-Moses et al. (2011) ``Day Avg."  }
\end{figure*}

\begin{figure*}[h]
\begin{center}
\includegraphics[width=0.75\textwidth, angle=0]{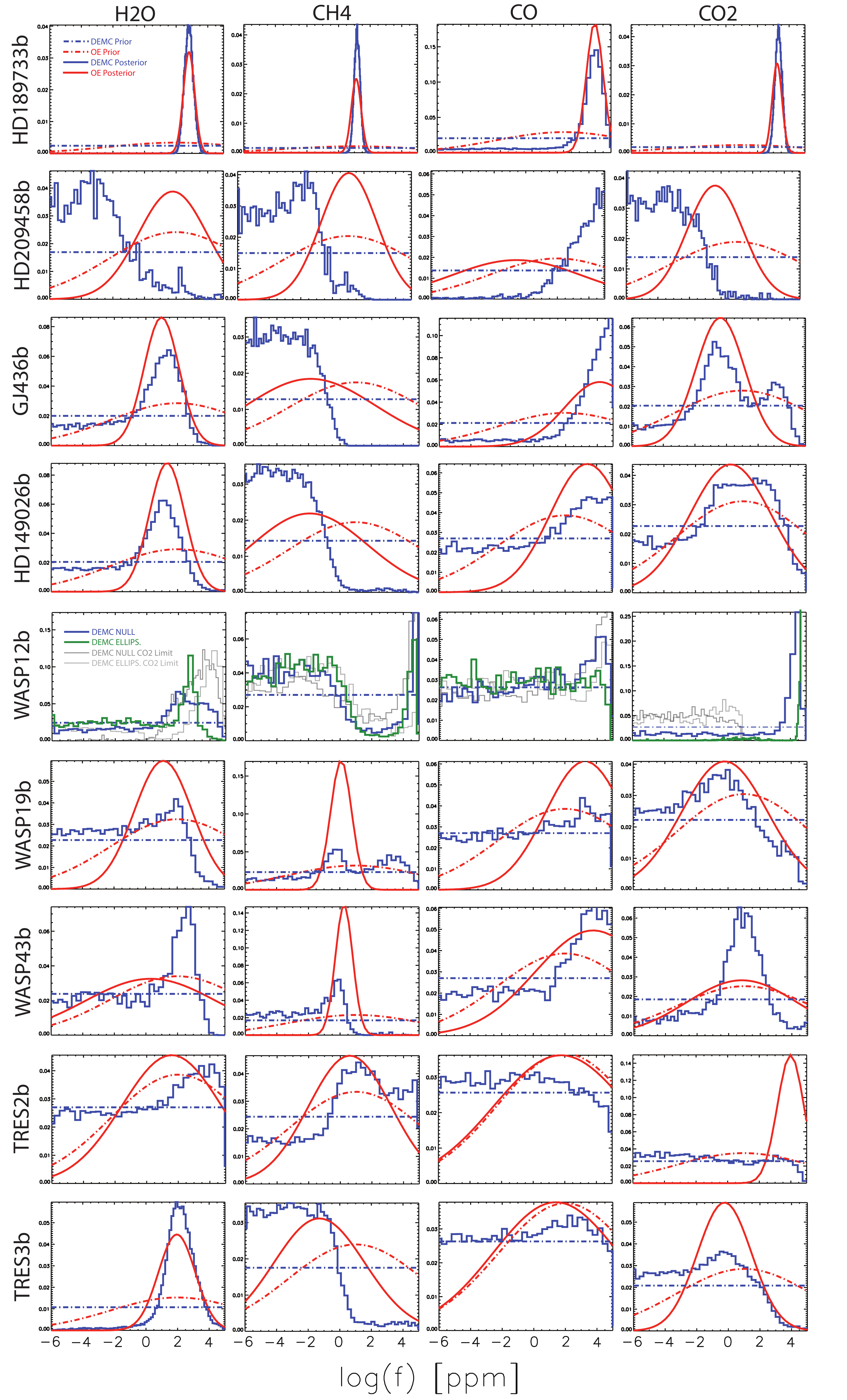}
\end{center}
     \caption[Mixing Ratio Vertical Profiles]{ \label{fig:Figure3} Histograms of the marginalized posterior probability distribution for each of the retrieved gases (columns) for each planet (rows).  The Gaussian probability distributions derived from the optimal estimation retrievals are shown in red and differential evolution Markov chain Monte Carlo results are in blue.  The priors for DEMC and OE are shown as the blue and red dot-dashed curves.  The y-axis is the normalized probability density for each gas with arbitrary units.  For WASP-12b we have included the DEMC histograms resulting from both the ``null" and ``ellipsoidal" variations in the 4.5 $\mu$m data (blue and red, respectively), and from imposing an upper limit of 10$^{-5}$ to the CO$_2$ abundance (gray, see \S\ref{sec:WASP12b}). }
\end{figure*}

\begin{figure*}[h]
\begin{center}
\includegraphics[width=0.75\textwidth, angle=0]{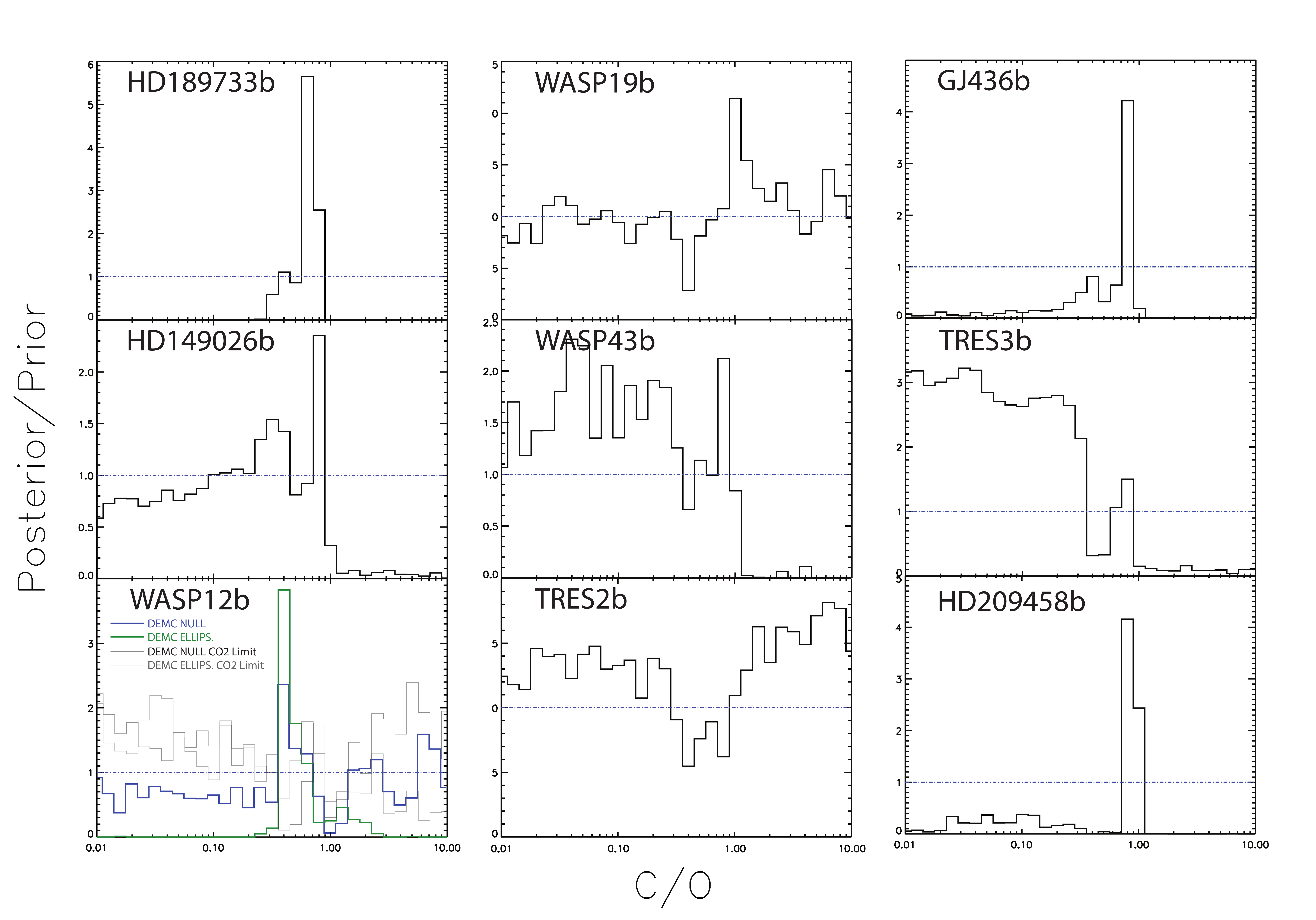}
\end{center}
     \caption[Mixing Ratio Vertical Profiles]{ \label{fig:Figure4} Resulting prior-normaized C/O ratio probability distributions for each planet.  These distributions are derived by dividing the double peaked prior described in Paper I into the C/O distributions that result from the posterior gas distributions.  Although this has no statistical meaning, it is a useful way to visualize how the data contributes to our knowledge of the C/O.  The horizontal blue dot-dashed line is the curve resulting from the C/O prior in Part I divided by itself.  See \S \ref{sec:Analysis} for more details.  For WASP-12b we have included the DEMC histograms resulting from both the ``null" and ``ellipsoidal" variations in the 4.5 $\mu$m data (blue and red, respectively), and from imposing an upper limit of 10$^{-5}$ to the CO$_2$ abundance (gray, see \S\ref{sec:WASP12b}).   }
\end{figure*}

In the future, spectroscopic observations of these planets with the HST WFC3 instrument offer a promising avenue for further constraining their C to O ratios and refining our knowledge of their atmospheres.  As can be seen in Figure \ref{fig:Figure1}, HD209458b, WASP-19b, WASP-43b, TrES-2b, and TrES-3b all have widely divergent spectra between 1 and 2 $\mu$m.  Comparing these spectral spreads to that for WASP-12b, we find that the WFC3 data reduces the overall uncertainty in the spectra between 1.1 and 1.8 $\mu$m.   Looking at Figure \ref{fig:Figure2}, we can see that the WFC3 data can constrain temperatures in the deep atmosphere to a higher precision than those planets without the WFC3 data. If comparable signal-to-noise could be obtained for the other planets using WFC3, we could expect similar retrieved temperature precisions in the 1-10 bar regions of the atmosphere.   WFC3 measurements of HD189733b during secondary eclipse would also help to confirm or refute the high CO$_2$ abundance required to explain the NICMOS data.  In the longer term, higher resolution, high signal-to-noise instruments such as those planned for the James Webb Space Telescope or a future dedicated exoplanet atmosphere characterization mission will enable definitive determinations of the atmospheric properties of hot Jupiters.

\begin{table*}[h]
\centering
\caption{\label{tab:Table0} Planet and system parameters.  }
\begin{tabular}{ccccccc}
\hline
\hline
\cline{1-2}
Planet & R$_{p}$ (R$_{Jup}$)  &  R$_{*}$ (R$_{\odot}$) & T$_{*}$ (K) & a (A.U.) & log(g$_p$) (cm s$^{-2}$) & Ref.   \\
\hline
HD189733b	&1.138	&0.756	&5040	&0.03099	&3.34	&Torres et al. (2008)\\
HD209458b	&1.359	&1.155	&6065	&0.04707	&2.96	&Torres et al. (2008)\\
GJ436b  		&0.377      &0.464 	& 3350 	&0.02872 &3.11 	&Torres et al. (2008)\\
HD149026b	& 0.654	&  1.368	&  6160	&0.04313 & 3.13	&Torres et al. (2008); Carter et al. (2009)	\\
WASP-12b	& 1.79	&  1.57	&  6300	& 0.0229	& 3.0	0	&Hebb et al. (2009)\\
WASP-19b	& 1.386	&  0.990	&  5500	&0.01632	& 3.19	&Hebb et al. (2010); Hellier et al. (2011a)\\
WASP-43b	& 0.93	&  0.60	&  4400    & 0.0142	& 3.67	&Hellier et al. (2011b)\\
TrES-2b		&1.224	& 1.003	& 5850	&0.03558	& 3.30	&Torres et al. (2008)\\
TrES-3b		& 1.336	&  0.812	&  5650	& 0.02272	& 3.4	5	&Torres et al. (2008); Sozzetti et al. (2009)\\
\hline
\end{tabular}
\end{table*}


 \begin{table*}[h]
\centering
\caption{\label{tab:Table1} Sources of secondary eclipse data. The wavelengths of the channels are given in microns. }
\begin{tabular}{c  p{8cm}  p{8cm} }
\hline
\hline
\cline{1-2}
Planet & Spectroscopic Data Sources & Broadband Data Sources  \\
\hline
HD189733b  &  HST NICMOS\tablenotemark{a} (Swain et al. 2009), Spitzer IRS\tablenotemark{b} (Grillmair et al. 2007)  &  Spitzer IRAC\tablenotemark{c} 3.6, 4.5 (Knutson et al. 2012), 5.7, 7.8 (Agol et al. 2010) IRS 16,MIPS\tablenotemark{d}  24 (Charbonneau et al. 2008) \\
\\
\hline
HD209458b  & -   & Spitzer IRAC 3.6, 4.5, 5.7 (Knutson et al. 2008), 7.8 (H. Knutson et al. private com.), IRS 16 (Madhusudhan \& Seager 2009 via D. Deming private com),MIPS  24 (Crossfield et al. 2012a)  \\
\\
\hline
GJ436b  & -   & Spitzer IRAC 3.6, 4.5, 5.7, 7.8, IRS 16,MIPS  24 (Stevenson et al. 2010)  \\
\\
\hline
HD149026b  & -   & Spitzer IRAC 3.6, 4.5, 5.7, 7.8, IRS 16 (Stevenson et al. 2012)   \\
\\
\hline
WASP-12b  &  HST WFC3\tablenotemark{e} (Swain et al. 2013)  & CFHT WIC 1.66 (Croll et al. 2011), 2.15  (Zhao et al. 2012), Subaru NB 2.315 (Crossfield et al. 2012), Spitzer IRAC 3.6, 4.5\tablenotemark{f}, 5.7, 7.8 (Crossfield et al. 2012 )   \\
\\
\hline
WASP-19b  & -   &  VLT HAWK-I\tablenotemark{g}1.620 (Anderson et al. 2010), 2.095 (Gibson et al. 2010), Spitzer IRAC 3.6, 4.5, 5.7, 7.8 (Anderson et al. 2013) \\
\\
\hline
WAS43b & -    &  VLT HAWK-I 1.19 (Gillon et al. 2012), 1.620 (Wang et al. 2013), 2.1 (Gillon et al. 2012), Spitzer IRAC 3.6, 4.5 (Blecic et al. 2013) \\
\\
\hline
TrES-2b  &-    & CFHT WIC\tablenotemark{h} 2.15 (Croll et al. 2010a), Spitzer IRAC 3.6, 4.5, 5.7, 7.8 (O'Donovan et al. 2010)    \\
\\
\hline
TrES-3b  & -   & CFHT WIC 2.15 (Croll et al. 2010b), Spitzer IRAC 3.6, 4.5, 5.7, 7.8 (Fressin et al. 2010)   \\
\\
\hline
\end{tabular}
$^{a}$Hubble Space Telescope Near Infrared Camera and Multi-Object Spectrometer \\
$^{b}$Infrared Spectrometer\\
$^{c}$nfrared Array Camera  \\
$^{d}$Multiband Imaging Photometer for Spitzer\\
$^{e}$Wide Field Camera 3\\
$^{f}$We use both the null and ellipsoidal cases from Cowan et al. 2012 as corrected by Crossfield et al. 2012\\
$^{g}$Very Large Telescope High Acuity Wide field K-band Imager\\
$^{h}$Canada-France-Hawaii Telescope Wide-field Infrared Camera\\
\end{table*}

\begin{table*}[h]
\caption{\label{tab:Table2} Summary of the abundance retrieval results for each planet compared with the literature.\tablenotemark{a}     }
\resizebox{\textwidth}{!}{%
\begin{tabular}{ccccccc}
\hline
\hline
\cline{1-2}
Planet & Source & H$_2$O\tablenotemark{b} & CH$_4$\tablenotemark{b} & CO\tablenotemark{b} & CO$_2$\tablenotemark{b} & C/O\tablenotemark{c}\\
\hline
HD189733b      & Best Fit ($\chi^2_{best}/N=$2.27) & $8.86\times10^{-4}$ &	$19.8\times10^{-6}$ &	$182\times10^{-4}$ &	$2.87\times10^{-3}$ &	0.85\\
			& 68$\%$ Interval & $[3.38-12.8]\times10^{-4}$ & $[9.97-23.9]\times10^{-6}$ & $[0.275-307]\times10^{-4}$ & $[1.29-3.73]\times10^{-3}$ &  0.47 - 0.90\\
			& Madhusudhan \& Seager 2009 &   $[0.1-10]\times10^{-4}$ &$<$few$\times10^{-6}$ & $[1-100]\times10^{-4}$ &  $\sim70\times10^{-3}$  & 0.007 - 1	\\	
			& Swain et al. 2009 & $[0.1-1]\times10^{-4}$ & $<10\times10^{-6}$  &$[1-3]\times10^{-4}$  & $[0.0001-0.001]\times10^{-3}$ &  0.5 - 1 \\
			& Lee et al. 2012 &   $[0.3-100]\times10^{-4}$	 &$<100\times10^{-6}$ & -	& $[0.15-30]\times10^{-3}$ & 0.30 - 1\\
			& Line et al. 2012 & $[0.5-3]\times10^{-4}$  & $<10000\times10^{-6}$  & $[36-360]\times10^{-4}$ & $[1.7-6.7]\times10^{-3}$ & -\\
\\
\hline
HD209458b      & Best Fit ($\chi^2_{best}/N=$0.31) & $4.95\times10^{-9}$ &	$3.04\times10^{-8}$ &	$5.91\times10^{-1}$ &	$5.22\times10^{-9}$ &	1.00\\
			& 68$\%$ Interval & $[0.0093-108]\times10^{-9}$ & $[0.0011-9.24]\times10^{-8}$ & $[0.0048-3.69]\times10^{-1}$ & $[0.0062-27.3]\times10^{-9}$ &  1.00 - 1.00\\
			& Madhusudhan \& Seager 2009 &   $[10-10000]\times10^{-9}$ &$[4-3000000] \times10^{-8}$ & $>0.004\times10^{-1}$ &  $[4-70]\times10^{-9}$  & $\ge$1	\\	
			& Swain et al. 2009 &   $[800-100000]\times10^{-9}$ &$[2000-20000] \times10^{-8}$ & - &  $[1000-10000]\times10^{-9}$  & -	\\	

\\
\hline
GJ436b		& Best Fit ($\chi^2_{best}/N=$1.78) 	&	$5.56\times10^{-6}$ &	$5.65\times10^{-9}$ &	$34.5\times10^{-3}$ &	$3.18\times10^{-7}$ &	  1.0\\
			& 68$\%$ Interval &  $[0.00071-105 ]\times10^{-6}$  &   $[0.0067-24.5]\times10^{-9}$ & $[0.016-66.4]\times10^{-3}$ &   $[0.0411-8720]\times10^{-7}$ & 0.50 - 1.0\\
			& Madhusudhan \& Seager 2011 & 	$<100\times10^{-6}$  &	$[100-1000 ]\times10^{-9}$ &	$>1\times10^{-3}$	&	 $[10-1000 ]\times10^{-7}$ &	0.5 - 1.0 \\
			& Stevenson et al. 2010 &	$3\times10^{-6}$ &$ 100\times10^{-9}$ &	$0.7\times10^{-3}$ &   $1\times10^{-7}$  &   - 	\\
\\			
\hline
HD149026b	& Best Fit ($\chi^2_{best}/N=$0.23) &	$172\times10^{-7}$ &	$0.032\times10^{-10}$ &	$21\times10^{-6}$  &	$19\times10^{-7}$  &	  0.55 \\
			& 68$\%$ Interval & $[0.0034-890]\times10^{-7}$  & $[0.072-290]\times10^{-10}$ &  $[0.00024-11400]\times10^{-6}$	&   $[0.0042-7170]\times10^{-7}$ &  0.45 - 1.0   \\
\\
\hline
WASP-12b	
			& Best Fit-Null ($\chi^2_{best}/N=$1.74) & $0.0082\times10^{-6}$ &	$0.024\times10^{-7}$ & $2210\times10^{-6}$ &$56400\times10^{-6}$   &	  0.51 \\
			& 68$\%$ Interval-Null &  $[0.0016-5430 ]\times10^{-6}$  &  $[0.00017-8270]\times10^{-7}$	 & $[0.00018-10400]\times10^{-6}$ &  $[0.0074-48400]\times10^{-6}$ &  0.30 - 1.00 \\
			& Best Fit-Ellipsoidal ($\chi^2_{best}/N=$1.71) & $512\times10^{-6}$ &	$0.002\times10^{-7}$ & $2170\times10^{-6}$ &$107000\times10^{-6}$   &	  0.50 \\
			& 68$\%$ Interval-Ellipsoidal &  $[0.00011-863 ]\times10^{-6}$  &  $[0.00028-75.3]\times10^{-7}$	 & $[0.000051-1460]\times10^{-6}$ &  $[51100-146000]\times10^{-6}$ &  0.11-0.22 \\
			& Madhusudhan et al. 2011 & 	$[0.00005-6.00]\times10^{-6}$ &	$[40-8000 ]\times10^{-7}$ &	$[30-3000]\times10^{-6}$	&	 $[0.2-7]\times10^{-6}$	&	         $>$1.0  \\
			& Swain et al. 2013 &	$[0.000000047-28000]\times10^{-6}$ &$ [0.9-60.12]\times10^{-7}$ &	$[0.0000011-567000]\times10^{-6}$	 &   $[0.007 - 2400]\times10^{-6}$    &   0.3 	\\
\\
\hline
WASP-19b	& Best Fit ($\chi^2_{best}/N=$0.032)  &	$130\times10^{-7}$ &	$1.21\times10^{-6}$ & $1900\times10^{-6}$ & $62\times10^{-8}$ &	 0.99\\
			& 68$\%$ Interval &  $[0.00046-1080 ]\times10^{-7}$  &  $[0.0015-3890]\times10^{-6}$ &$[0.00013-5110]\times10^{-6}$ &  $[0.0053-5950]\times10^{-8}$&  0.26 - 6.33\\
			& Madhusudhan 2012\tablenotemark{d} & 	$[1000/20]\times10^{-7}$ &	$[0.0006/0.5]\times10^{-6}$ &	$[600/500]\times10^{-6}$ & $[0.06/0.001]\times10^{-7}$ &	0.4/1.1  \\
\\
\hline
WASP-43b	& Best Fit ($\chi^2_{best}/N=$0.52)  	&	$1.98\times10^{-6}$ &	$1780\times10^{-9}$ &	$6090\times10^{-6}$  &	$67.9\times10^{-7}$ 	&	  1.00\\
			& 68$\%$ Interval & $[0.0002-706 ]\times10^{-6}$  &  $[0.026-1050]\times10^{-9}$  & $[0.00052-17000]\times10^{-6}$  &  $[0.082-1310]\times10^{-7}$  &  0.132-1.00 	\\	
\\
\hline
TrES-2b		& Best Fit ($\chi^2_{best}/N=$0.60) 	&	$132\times10^{-6}$ &	$24.6\times10^{-6}$ &	$22.4\times10^{-7}$  &	$0.00010\times10^{-7}$ &	  0.20 \\
			& 68$\%$ Interval  & $[0.00013-6300 ]\times10^{-6}$ &$[0.0011-6340]\times10^{-6}$ &$[0.00055-11300]\times10^{-7}$ & $[0.00035-8760]\times10^{-7}$ &  0.021-8.25\\
\\
\hline
TrES-3b		&  Best Fit ($\chi^2_{best}/N=$0.067) &	$0.90\times10^{-4}$ &	$50.4\times10^{-9}$ &	$247\times10^{-7}$ &	$57\times10^{-8}$  &	  0.22 \\
			& 68$\%$ Interval & $[0.13-12.5 ]\times10^{-4}$&$[0.026-257]\times10^{-9}$ &$[0.00076-22500]\times10^{-7}$ & $[0.0034-1880]\times10^{-8}$& 0.0004-0.97\\
\\
\hline
\end{tabular}}
$^{a}$We present the best fit mixing ratios and the resulting C/O ratio along with their 68$\%$ confidence intervals.  The best fit is defined as the fit that produces the minimum cost function value, $\chi_{best}^2$ from equation \ref{eq:cost_func}.  We quote the reduced cost-fucntion value as the cost function value divided by the number of data points, $N$.   Note that the best-fit values for some of the parameters can fall outside of the 68\% confidence interval.\\
$^{b}$Since we place upper and lower limits on the DEMC prior, these results must be interpreted in the context of the prior.  The 68$\%$ confidence interval resulting from the flat priors alone would give values that range from $8.15\times10^{-11} - 1.51\times10^{-2}$, or $\sim$8 orders of magnitude.  One should use caution when interpreting uncertainties that approach these levels, which is indicative of little information gain from the data. \\
$^{c}$The C/O ratios resulting from the prior would span $0.051 - 14.5$, or  $\sim$ 3 orders of magnitude. \\
$^{d}$We show both the oxygen rich planet (ORP) values and the carbon rich planet (CRP) values quoted in his investigation [ORP/CRP] \\

\end{table*}

\section{Acknowledgements}		
We thank Jonathan Fortney, Ian Crossfield, Xi Zhang, Nikku Madahusudhan, Caroline Morely, and Jonathan Frain for insightful conversations.  We also thank the anonymous referee for their insightful comments that made this paper stronger.  This research was supported in part by an NAI Virtual Planetary Laboratory grant from the University of Washington to the Jet Propulsion Laboratory and California Institute of Technology.

\end{document}